\documentclass[12pt]{article}
\usepackage{epsfig,amsfonts,amssymb,amsbsy,amsbsy,array}
\def\cvp{\raise 2pt\hbox{,}}

\def\tr{\mathop{\rm tr}\nolimits}
\def\im{\mathop{\rm Im}\nolimits}
\def\re{\mathop{\rm Re}\nolimits}

\def\d{{\rm d}}
\def\suN{{\rm SU}(N)}
\def\uN{{\rm U}(N)}

\def\wl{W_{\rm low}}\def\wq{W_{\rm q}}
\def\wt{W_{\rm tree}}\def\weff{W_{\rm eff}}\def\ww{{\cal W}}
\def\Ld{\Lambda_{\rm d}}

\def\C{{\rm C}}\def\F{{\cal F}}

\def\plb#1#2#3{{\it Phys.\ Lett.\ }{\bf B #1} (#2) #3}
\def\npb#1#2#3{{\it Nucl.\ Phys.\ }{\bf B #1} (#2) #3}
\def\npps#1#2#3{{\it Nucl.\ Phys.\ Proc.\ Suppl.\ }{#1} (#2) #3}
\def\prl#1#2#3{{\it Phys.\ Rev.\ Lett.\ }{\bf #1} (#2) #3}
\def\jhep#1#2#3{{\it J. High Energy Phys.\ }{\bf #1} (#2) #3}
\def\prd#1#2#3{{\it Phys.\ Rev.\ }{\bf D #1} (#2) #3}

\def\atmp#1#2#3{{\it Adv.\ Theor.\ Math.\ Phys.\ }{\bf #1} (#2) #3}
\def\cmp#1#2#3{{\it Comm.\ Math.\ Phys.\ }{\bf #1} (#2) #3}
\def\pr#1#2#3{{\it Phys.\ Rep.\ }{\bf #1} (#2) #3}
\def\jmp#1#2#3{{\it J.\ Math.\ Phys.\ }{\bf #1} (#2) #3}
\def\ijmpa#1#2#3{{\it Int.\ J.\ Mod.\ Phys.\ }{\bf A #1} (#2) #3}
\def\mpla#1#2#3{{\it Mod.\ Phys.\ Lett.\ }{\bf A #1} (#2) #3}

\begin{document}
\def\L{\Lambda}
%
\pagestyle{empty}
{\parskip 0in

\hfill NEIP-02-007

\hfill LPTENS-02/50

\hfill hep-th/0210135}

\vfill
\begin{center}
{\LARGE On exact superpotentials in confining vacua}



\vspace{0.4in}

Frank F{\scshape errari}{\renewcommand{\thefootnote}{$\!\!\dagger$}
\footnote{On leave of absence from Centre 
National de la Recherche Scientifique, Laboratoire de Physique 
Th\'eorique de l'\'Ecole Normale Sup\'erieure, Paris, France.}}
\\
\medskip
{\it Institut de Physique, Universit\'e de Neuch\^atel\\
rue A.-L.~Br\'eguet 1, CH-2000 Neuch\^atel, Switzerland}\\
\smallskip
{\tt frank.ferrari@unine.ch}
\end{center}
\vfill\noindent
We consider the ${\cal N}=1$ super Yang-Mills theory with gauge group
$\uN$ or $\suN$ and one adjoint Higgs field with an arbitrary
polynomial superpotential.
We provide a purely field theoretic derivation of the
exact effective superpotential $W(S)$ for the glueball superfield
$S=-W^{a\alpha}W^{a}_{\alpha}/(32 N\pi^{2})$ in the confining vacua.
We show that the result matches with the Dijkgraaf-Vafa matrix model
proposal. The proof brings to light a deep relationship between
non-renormalization theorems first discussed by Intriligator, Leigh
and Seiberg, and the fact that $W(S)$ is given by a sum over planar
diagrams.

\vfill

\medskip
%
\begin{flushleft}
September 2002
\end{flushleft}
\newpage\pagestyle{plain}
\baselineskip 16pt
\setcounter{footnote}{0}

%
\section{Introduction}

In a recent insightful paper, Dijkgraaf and Vafa \cite{DV} have 
proposed a very simple recipe to calculate the exact quantum effective 
superpotential $W(S)$ for the glueball superfield
\begin{equation}
\label{Sdef}
S = -{\tr W^{\alpha}W_{\alpha}\over 16N\pi^{2}} = 
-{W^{a\alpha}W_{\alpha}^{a}\over 32N\pi^{2}}
\end{equation}
in the confining vacua of 
a large class of ${\cal N}=1$ supersymmetric Yang-Mills theories. The 
superpotential $W(S)$ contains highly non-trivial information about 
the non-perturbative dynamics of the theory. For example, 
it can be used to derive 
dynamical chiral symmetry breaking and calculate the tension of 
the associated domain walls. The work of Dijkgraaf and Vafa is the 
outcome of a long series of work on a large $N$ string theory 
duality, first proposed in \cite{GV} and further developed in \cite{CIV} 
and \cite{DV1}. The result is particularly useful because field 
theoretic derivations of exact glueball superpotentials in non-trivial 
cases have not appeared.

The purpose of the present paper is to remedy this situation, by 
providing a full field theoretic derivation in the archetypal 
example of the $\uN$ or $\suN$ theory with one adjoint Higgs 
supermultiplet $\Phi$ and a tree level superpotential of the general form
\begin{equation}
\label{Wtreedef}
\wt  = \sum_{p\geq 1}{g_{p}\over p} \tr\Phi^{p}= \sum_{p\geq 1}
g_{p} u_{p}\, .
\end{equation}
The lagrangian of the theory is
\begin{equation}
\label{lag}
{\cal L} = {N\over 4\pi}\im\tr\tau\left[\int\!\d^{2}\theta\,
W^{\alpha}W_{\alpha} + 2\int\!\d^{2}\theta\d^{2}\bar\theta\, 
\Phi^{\dagger} e^{2V}\Phi\right] + 
2N\re\!\int\!\d^{2}\theta\, \wt(\Phi)\, ,
\end{equation}
where $\tau$ is the complexified 't Hooft coupling constant. 
The classical glueball superpotential can be read off from (\ref{lag}), 
\begin{equation}
\label{wcl}
W_{\rm cl}(S) = 2 i\pi N \tau S\, .
\end{equation}
We will focus in Section 2 on the field theory calculation yielding
the exact quantum superpotential $W(S)$ in the confining vacua where
classically $\langle\phi\rangle =0$. Non-trivial field theoretic 
results on this theory can be found in \cite{CIV,CV}.
Our main tools are the
Seiberg-Witten solution for the ${\cal N}=2$ theory which is obtained
by turning off $\wt$ \cite{SW1,SW2}, and the Intriligator-Leigh-Seiberg
linearity principle \cite{ILS}. For the sake of clarity, and since
this is an important aspect of our work, we give a self-contained
account of this principle. We then analyse in Section 3 the
Dijkgraaf-Vafa proposal for our model. We use some matrix model
technology to put the solution in a simple form. We can then
straightforwardly compare the matrix model and the field theory
results, and we find perfect agreement. In Section 4 we recapitulate
our findings and add a few comments on future directions of research.

The original motivation for the present work was actually to use the
results in \cite{DV} to work out some new interesting physics along
the lines of preceding papers by the author \cite{F}.
These developments will appear in a separate publication \cite{fer}.

\section{Field theoretic analysis}
\subsection{The Intriligator-Leigh-Seiberg linearity principle}

The basic tools to analyse ${\cal N}=1$ supersymmetric gauge theories
were provided by Seiberg long ago \cite{sei1}. An excellent review
with a list of references is \cite{revN1}. A general procedure
consists in using holomorphy, symmetries, and various known
asymptotics and consistencies to derive the most general form of the
quantum 1PI superpotential for a set of fields $X_{r}$ as a
function of the complex mass scale $\L$ governing the one-loop running 
of the gauge coupling constant (we limit
the discussion for simplicity to the case where only one mass scale is
present). In a wealth of examples, particularly when there is no tree
level superpotential $\wt$, those constraints actually fix the
superpotential $\wq^{0}(X_{r},\L)$ uniquely. An interesting problem
is then to turn on
\begin{equation}
\label{suplin}
\wt = \sum_{r} g_{r} \tilde X_{r}\, ,
\end{equation}
where the $\tilde X_{r}$ form a subset of the $X_{r}$ which can be
expressed locally in terms of the elementary fields and can thus be
included consistently in the bare lagrangian. In some examples, the
use of the general constraints may still determine uniquely $\wq
(X_{r},\L,g_{r})$, but usually this is no longer the case.
Intriligator, Leigh and Seiberg (ILS for short) then proposed in
\cite{ILS} that the exact superpotential should be a simple linear
function in the bare couplings $g_{r}$,
\begin{equation}
\label{LP}
\wq (X_{r}, \L, g_{r}) = \wq^{0}(X_{r},\L) + \sum_{r} g_{r} \tilde
X_{r}\, .
\end{equation}
This states that the tree level superpotential for the $\tilde X_{r}$
is not renormalized, neither perturbatively (as is well-known) nor
non-perturbatively. Of course, that does not preclude a possible 
renormalization of the fields and the couplings taken individually. 
When such a renormalization does not occur perturbatively, we will 
also assume that it does not non-perturbatively.
The non-trivial renormalizations forbidden by the ILS principle are
not to be confused with possible vacuum independent ambiguities in the
definition of operators, as discussed for example in \cite{DH}.

A particularly important physical consequence is that, since there is
no renormalization at work, the fields $\tilde X_{r}$ can be
integrated out without loosing any information. Mathematically, this
comes from the linearity in the couplings $g_{r}$. This linearity
implies that the integrating out can be reversed by a simple
Legendre transform, a procedure called ``integrating in'' \cite{ken}.
In practice, one can then work with a superpotential for which all the
fields have been integrated out,
\begin{equation}
\label{wlowdef}
\wl (\L,g_{r}) = \wq (\langle X_{r}\rangle, \L,g_{r})\, .
\end{equation}
If necessary, the fields $\tilde X_{r}$ can then be integrated in by 
using the relation
\begin{equation}
\label{intin}
{\partial\wl\over\partial g_{r}} = \tilde X_{r}\, .
\end{equation}

To understand more concretely the significance of the ILS hypothesis,
let us consider the $\uN$ theory with lagrangian
(\ref{lag}). In that case, a standard set of fields
$X_{r}$ includes the monopole fields
$M_{m}$ and $\tilde M_{m}$, $1\leq m\leq N-1$, and the
$u_{p}=\tr\phi^{p}/p$, $1\leq p\leq N$. The fields $\tilde X_{r}$
correspond to the $u_{p}$s only. When $\wt\rightarrow 0$, the effective
superpotential is given by \cite{SW1}
\begin{equation}
\label{w01}
\wq^{0}(\tilde M_{m} M_{m},u_{p},\L) = \sqrt{2}\sum_{m=1}^{N-1}
\tilde M_{m} M_{m} A_{D,m}(u_{p},\L)\, ,
\end{equation}
where the $A_{D,m}$ are 
dual ${\cal N}=2$ ${\rm U}(1)$ 
vector multiplets scalars, which are known functions of the moduli
$u_{p}$ and scale $\L$ \cite{SW1,SW2}. 
Let us now add the tree level superpotential 
(\ref{Wtreedef}). The ILS hypothesis implies that the new exact 
effective superpotential is
\begin{equation}
\label{weff2}
\wq = \sqrt{2}\sum_{m=1}^{N-1}
\tilde M_{m} M_{m} A_{D,m}(u_{p},\L) + \sum_{p\geq 1} g_{p}u_{p}\, .
\end{equation}
Because we have ${\cal N}=2$ supersymmetry when $\wt =0$, and with the
normalization for the kinetic term of $\Phi$ in (\ref{lag}), there is
no renormalization for the individual fields $u_{p}$. The $u_{p}$s
appearing in (\ref{weff2}) are thus {\it the same as the UV operators
$u_{p}$ of the $\wt =0$ theory.} These facts can actually be proven
from symmetry and regularity arguments for a purely quadratic tree
level superpotential \cite{SW1}. However, in general, the same
symmetry and regularity arguments are no longer able to fully
determine $\weff$. For example, if $g_{2}$ and $g_{3}$ are turned on,
one can form the parameter $r=g_{3}^{2}\L^{2}/g_{2}^{2}$ which is
neutral under all the symmetries of the problem. Non-trivial
renormalizations involving arbitrary functions of this parameter are
possible in principle. The ILS hypothesis implies that those
renormalizations do not occur.

An important fact is that the gauge kinetic term 
can be viewed as an $F$-term with a superpotential given by 
(\ref{wcl}). Of course, this is also a $D$-term, and is thus renormalized. 
By working with renormalized matter fields
in the path integral measure (the kinetic terms then include
the non-trivial wave function renormalization factors Z), the 
renormalization of the gauge kinetic term is exhausted at one loop,
\begin{equation}
\label{tauren}
N\tau = {iNb\over\pi}\ln {\L_{0}\over\L}\,\cdotp
\end{equation}
The scale $\L_{0}$ is the ultraviolet cutoff, $b$ a number of order
one given by the one-loop $\beta$ function ($b=1$ for (\ref{lag})),
and $\L$ a complexified mass scale. This is the famous perturbative
non-renormalization theorem for the holomorphic Wilson gauge coupling
\cite{SV}. Extending this result to the non-perturbative realm by
applying the ILS principle, we find that the exact 1PI superpotential
including the glueball superfield (\ref{Sdef}) is of the general form
\begin{equation}
\label{WeffS}
W_{\rm 1PI} (X_{r},S,\L,g_{r}) = W^{0}(X_{r},S) + S\ln
\L^{2Nb} + \sum_{r}g_{r}\tilde X_{r}\, ,
\end{equation}
where we have indicated the dependence in all the variables explicitly. A
most crucial point for our purposes
is that the ``universal,'' coupling independent,
superpotential $W^{0}$ can be obtained, without making any further
assumption, from $\wq^{0}$ in (\ref{LP}) by integrating in. This is
because the full $\L$ dependence is always taken into account in
(\ref{LP}), even though the $S$ field does not appear (technically,
there is no $W_{I}$ in the notations of \cite{ken}, equation 2.2).
Explicitly, one solves
\begin{equation}
\label{Seq}
{\partial\wq^{0}\over\partial\ln\L^{2bN}} = S
\end{equation}
to express $\L$ as a function $f(X_{r},S)$, and writes
\begin{equation}
\label{W0}
W^{0}(X_{r},S) = -S\ln f(X_{r},S)^{2Nb} + 
\wq^{0}\left(\strut X_{r},\L = f(X_{r},S)\right)\, .
\end{equation}

Another subtle issue, that must be kept in mind when dealing with
general ${\cal N}=1$ models, is that the fields $S$ or $X_{r}$, scale
$\L$, and couplings $g_{r}$, that enter formulas like (\ref{WeffS}) or
the Dijkgraaf-Vafa superpotential $W(S)$ \cite{DV}, are not physical,
RG invariant quantities. In particular, they do not have a finite limit
when the UV cut-off $\L_{0}$ is taken to infinity. (Of course the
effective superpotentials themselves are physical and RG invariant).
The reason is that, to maintain holomorphicity, one must work with the
Wilson gauge coupling (\ref{tauren}) and renormalized fields and
couplings that are related to the physical quantities through
non-holomorphic $Z$ factors (a useful discussion of this problem, with
a complete list of references, can be found in \cite{RR}). 
For the model (\ref{lag}) studied in the present paper,
as well as for any model that can be viewed as a
perturbation of an ${\cal N}=2$ theory, the $Z$ factors are trivial
and thus this aspect is irrelevant. In particular, $\L$ 
corresponds in that case to a physical dynamically generated mass scale.

\subsection{Exact superpotentials}

We could now go on and, from the superpotential (\ref{weff2}), obtain
the universal superpotential $W^{0}(\tilde M_{m}M_{m},u_{p},S)$. This
would amount to calculating the Legendre transform of the periods
$A_{D,m}$. An elegant formula for $\partial
A_{D,m}/\partial\ln\L^{2N}$ can be found by using the exact RG
equations derived in \cite{RGN2}, but it is not clear how to find a
useful expression for $W^{0}$. Since our purpose is mainly to compute
the superpotential $W(S)$ for $S$ alone, we will first integrate out
the monopole fields from (\ref{weff2}), and then integrate in $S$. 
This is of course strictly equivalent to integrating in $S$ first and 
then integrating out the monopoles. We thus have to solve the 
equations $\partial\wq/\partial (\tilde M_{m}M_{m}) =0$, which yield
\begin{equation}
\label{ADeq}
A_{D,m}(u_{p},\L) = 0\, ,\quad 1\leq m\leq N-1\, .
\end{equation}
As explained in \cite{SW1,SW2}, the $A_{D,m}$ are given by integrals
of a known differential form over cycles of the genus $N-1$ Riemann
surface defined by the equation
\begin{equation}
\label{SWsol}
y^{2} = P(x) = \prod_{k=1}^{N} (x-x_{k})^{2} - 4\L^{2N}\, ,
\end{equation}
where the $x_{k}$s are such that
\begin{equation}
\label{Udef}
u_{p} = {1\over p}\sum_{k=1}^{N}x_{k}^{p}\, .
\end{equation}
We have normalized the scale $\L$ in (\ref{SWsol}) to obtain a simple 
match with the matrix model in the next Section.
The $A_{D,m}$ are zero when the corresponding cycles vanish, and this 
yields a factorization constraint on the polynomial $P(x)$. This 
constraint was solved for the case of $\suN$ in \cite{DS} with the help of 
Chebyshev polynomials. There are $N$ solutions, corresponding to the $N$ 
vacua of our ${\cal N}=1$ theory. We will generally use the solution
\begin{equation}
\label{DSsol}
x_{k} = 2\L \cos {\pi (k-1/2)\over N} \quad\Longleftrightarrow\quad
pu_{p}= \left\{ \matrix{\displaystyle 0&\!\!{\rm if\ }p\ {\rm is\ odd,}\cr
\displaystyle N\L^{p} {\rm C}_{p}^{p/2}
& {\rm if\ }p\ {\rm is\ even,}}\right.
\end{equation}
with the understanding that the other vacua are obtained
by $2\pi$ shifts of the $\theta$ angle, $\L^{2}\rightarrow\L^{2}e^{2i\pi 
k/N}$, $1\leq k\leq N-1$. The $\C_{n}^{p}$ are the binomial 
coefficients.
The case of $\uN$ can be boiled down to the case 
of $\suN$ by shifting the variables $x_{k}\rightarrow x_{k} +u_{1}/N$.
The most general solution to (\ref{ADeq}) is then straightforwardly 
obtained,
\begin{equation}
\label{sol}
u_{p} = U_{p}(z,\L^{2})= {N\over p}\sum_{q=0}^{[p/2]}\C_{p}^{2q}
\C_{2q}^{q} \L^{2q} z^{p-2q}\, ,
\end{equation}
where the variable $z$ is defined to be
\begin{equation}
\label{zdef}
z = u_{1}/N\, .
\end{equation}
The exact superpotential for the field $z$
can then be written down explicitly by replacing the solution 
(\ref{sol}) in (\ref{weff2}),
\begin{equation}
\label{Weffz}
\ww (z,\L^{2},g_{p}) = \sum_{p\geq 1} g_{p} U_{p}(z,\L^{2})\, .
\end{equation}
Integrating in $S$ then yields the effective superpotential $\weff$ 
for $z$ and $S$,
\begin{equation}
\label{ww}
\weff (z,S,\L^{2},g_{p}) =S\ln\L^{2N}+Ng_{1}z - S\ln \Delta (z,S,g_{p\geq 
2})^{N} + \sum_{p\geq 2} g_{p} U_{p}(z,\L^{2}=\Delta)\, ,
\end{equation}
where $\Delta (z,S,g_{p\geq 2})$ satisfies the condition
\begin{equation}
\label{Bc}
\L^{2}\partial_{\L^{2}}\ww (z,\L^{2}=\Delta ,g_{p}) =N S,
\end{equation}
together with the requirement that in the classical limit 
$S\rightarrow 0$, $\Delta\rightarrow 0$.

The superpotential $\weff$ is very 
convenient to use. For example, in the case of a cubic tree level
superpotential, whose physics is discussed in details in \cite{fer},
it takes the form
\begin{equation}
\label{ww3}
\weff^{\rm cubic}
(z,S,\L^{2},g_{p}) = N\Bigl( g_{1}z + {1\over 2}g_{2}z^{2} + {1\over 3} 
g_{3}z^{3}\Bigr) + S\ln \left({e\L^{2} (g_{2}+2g_{3}z)\over 
S}\right)^{N}\,\cdotp
\end{equation}
The superpotential (\ref{ww3}) can be used to 
describe all the vacua, because only $\L^{2N}$ enters the formula.

The derivatives of $\weff$ take simple forms,
\begin{eqnarray}
&& \partial_{S}\weff (z,S,\L^{2},g_{p}) = - \ln 
\left( \Delta (z,S,g_{p\geq 2})/\L^{2}\right)^{N} ,\label{D1}\\
&& \partial_{z}\weff (z,S,\L^{2},g_{p}) = \partial_{z} 
\ww \left(z,\L^{2}=\Delta (z,S,g_{p\geq 2}),g_{p}\right)\, .\label{D2}
\end{eqnarray}
We can use the condition
\begin{equation}
\label{dzw}
\partial_{z}\weff (z,S,\L^{2},g_{p}) = 0
\end{equation}
to integrate out $z$ and obtain the 
superpotential for the glueball superfield only. This yields
\begin{equation}
\label{Wfield}
W(S,\L^{2},g_{p}) = -S\ln \left( \Delta /\L^{2}\right)^{N} +
\sum_{p\geq 1} g_{p}U_{p}(z,\Delta )\, ,
\end{equation}
where the polynomials $U_{p}$ are defined in (\ref{sol}), and $\Delta$ and 
$z$ are expressed in terms of $S$ by using the two conditions 
(\ref{Bc}) and (\ref{dzw}). Explicitly, those conditions are
\begin{eqnarray}
&& S = \sum_{p\geq 2} 
g_{p}\sum_{q=1}^{[p/2]}{q\over p}\C_{p}^{2q}\C_{2q}^{q} z^{p-2q}
\Delta^{q}\, ,\label{Acex}\\
&& 0 = \sum_{p\geq 1} 
g_{p}\sum_{q=0}^{[(p-1)/2]}{p-2q\over p}\C_{p}^{2q}\C_{2q}^{q}
z^{p-2q-1} \Delta^{q}\, .\label{Bcex}
\end{eqnarray}
To compare with the matrix model result discussed in the next Section, 
it is convenient to use the derivative of $W$. From (\ref{D1}) we 
deduce the fundamental field theory formula
\begin{equation}
\label{dW}
\partial_{S}W(S,\L^{2},g_{p}) = -\ln\left(\Delta /\L^{2}\right)^{N}\, .
\end{equation}
Note that the full dependence in $\L$ is explicit in (\ref{dW}), since the
equations (\ref{Acex}) and (\ref{Bcex}) that determine $\Delta$ are
independent of $\L$. The linearity in $\ln\L^{2N}$ is of course a direct
consequence of the ILS principle. Equation (\ref{dW}) shows that to
calculate $\langle S\rangle$, one must simply set $\Delta=\L^{2}$ in
(\ref{Acex}) and (\ref{Bcex}) and solve the algebraic equations so
obtained.

The formulas simplify when $\wt$ is an even function, because  
the solution to (\ref{Bcex}) is then simply $z=0$, and thus (\ref{Wfield}) 
and (\ref{Acex}) reduce to
\begin{eqnarray}
&& W(S,\L^{2},g_{2p}) = -S\ln \left( \Delta /\L^{2}\right)^{N} +
N\sum_{p\geq 1}{1\over 2p} g_{2p}\C_{2p}^{p}\Delta^{p}\, ,\label{even1}\\
&& S = {1\over 2}\sum_{p\geq 1} g_{2p} \C_{2p}^{p}\Delta^{p}\,
.\label{even2}
\end{eqnarray}
The preceding equations also give the solution of the $\suN$ theory for 
an {\it arbitrary} tree level superpotential. This is proven by 
treating the coupling $g_{1}$ as a Lagrange multiplier in (\ref{Wfield}),
which automatically sets $z=0$.

\section{Matrix model analysis}
\subsection{The Dijkgraaf-Vafa proposal}

Dijkgraaf and Vafa have conjectured in \cite{DV}
that the superpotential $W(S)$
can actually be computed by summing the zero momentum
{\it planar} diagrams of the ${\cal N}=1$ theory under consideration. 
In our case, their ans\"atz for the $\uN$ theory
is simply an holomorphic integral over $n\times n$ complex matrices $\phi$,
\begin{equation}
\label{matDV}
\exp\left( n^{2}\F /S^{2}\right) = \int_{\rm planar}
\hskip -.5cm\d^{n^{2}}(\phi/\L)\,
\exp\Bigl[ -{n\over S}\,\wt (\phi,g_{p})\Bigr]\, ,
\end{equation}
from which the superpotential can be deduced,
\begin{equation}
\label{WDV}
W(S,\L^{2},g_{p}) = -N\partial_{S}\F (S,g_{p})\, .
\end{equation}
It is convenient to introduce the dummy variable $n$ (which
is not to be confused with the number of color $N$) because the planar 
diagrams can be extracted by taking the $n\rightarrow\infty$ limit. 
The full $N$ dependence of the superpotential is then given explicitly 
in (\ref{WDV}). Strictly speaking, the integral (\ref{matDV}) involves
complex matrices and couplings $g_{p}$, but the calculation is the same as
for hermitian matrices and real couplings. There is no ambiguity in the 
analytic continuation because we restrict ourselves to
planar diagrams. This implies that the standard matrix model 
techniques, which are reviewed for example in \cite{matrev}, do apply.
For the $\suN$ gauge theory, the integral (\ref{matDV})
must be restricted to traceless matrices, or equivalently one must
treat $g_{1}$ as a Lagrange multiplier. There is no difference between 
the $\uN$ and $\suN$ theory when the function $\wt$ is even, because 
the ${\rm U}(1)$ part of $\phi$ in the $\uN$ theory has then zero vev and 
its couplings are subleading. In general, however, the
$\uN$ and $\suN$ theories are very different \cite{fer}. This is perfectly 
consistent with the field theory results discussed at the end of the 
preceding Section.

It is convenient to work with dimensionless variables and to write
\begin{equation}
\label{DVint2}
\exp\left( n^{2}\F /S^{2}\right) = \int_{\rm planar}
\hskip -.5cm\d^{n^{2}}\varphi\,
\exp\Bigl[ -{n\over\sigma}\tr V(\varphi,\lambda_{p})\Bigr]\, ,
\end{equation}
where
\begin{equation}
\label{dimless}
\sigma = S/\Ld^{3}\, ,\quad
V(\varphi,\lambda_{p}) = \sum_{p\geq 1} {\lambda_{p}\over p}\, 
\varphi^{p}\, , \quad \lambda_{p} = 
{g_{p}\Ld^{3(p-2)/2}\over g_{2}^{p/2}}\,\cdotp
\end{equation}
The scale
\begin{equation}
\label{Lddef}
\Ld^{3} = g_{2}\L^{2}
\end{equation}
is the dynamically generated scale of 
the low energy pure ${\cal N}=1$ super Yang-Mills theory.

\subsection{Matrix model technology}

The method to calculate explicitly (\ref{DVint2}) has been known for a long
time \cite{BIPZ,matrev}. The eigenvalues of $\phi$, which are all zero
classically in the vacua we consider, are described in the planar limit by
a continuous distribution $\rho(\varphi ,\sigma)$ with support
on a finite interval $[a,b]$ containing zero (we will no longer indicate 
explicitly the dependence in the couplings $\lambda_{p}$).
The solution is easily expressed in terms of $\rho$,
\begin{equation}
\label{Fform}
{{\cal F}\over \Ld^{6}} = -\sigma\int\!\d\varphi\, 
\rho(\varphi ,\sigma)V(\varphi) + \sigma^{2}\int\!\d\varphi\d\psi\, 
\rho(\varphi ,\sigma)\rho(\psi ,\sigma)\ln |\varphi -\psi |\, .
\end{equation}
For our purposes, the most useful way to present the solution for $\rho$, 
$a$ and $b$ is in terms of the following three equations (see
\cite{matrev} for a derivation),
\begin{eqnarray}
&& \partial_{\sigma} \left(\strut\sigma\rho(\varphi,\sigma)\right) = 
{1\over\pi\sqrt{(\varphi -a)(b-\varphi)}}\, \cvp \label{rhoeq}\\
&& \sigma = \int_{a}^{b} {\d\varphi\over 2\pi}\,{\varphi V'(\varphi)\over
\sqrt{(\varphi -a)(b-\varphi)}}\, \cvp\label{m1}\\
&& 0 = \int_{a}^{b}\d\varphi\, { V'(\varphi)\over
\sqrt{(\varphi -a)(b-\varphi)}}\, \cdotp\label{m2}
\end{eqnarray}
The formulas (\ref{m1}) and (\ref{m2}) give two algebraic equations that 
determine $a$ and $b$ as a function of $\sigma$ and the couplings,
and (\ref{rhoeq}) may then be used to deduce $\rho$.
The superpotential (\ref{WDV}) is related to the derivative of
$\cal F$. To make this calculation,\footnote{This calculation is
undoubtedly known to experts in the field of matrix 
models, since the result (with a minor misprint) appears for example
in \cite{eynard}, but I have been unable to find a derivation in the 
literature.} we will need two identities,
\begin{eqnarray}
&& \int_{a}^{b} {\d\varphi\over\pi}\, {\ln |\varphi -\psi|\over
\sqrt{(\varphi -a)(b-\varphi)}} = \ln {b-a\over 4}\quad {\rm for\ any}\ 
\psi\in [a,b]\, ,\label{i1}\\
 \partial_{\sigma}\hskip -.7cm&& \int_{a}^{b} {\d\varphi\over\pi}\, 
{ V(\varphi)\over\sqrt{(\varphi -a)(b-\varphi)}} =
2\sigma\partial_{\sigma}\ln (b-a)\, ,\label{i2}
\end{eqnarray}
that we derive in the Appendix.
By using (\ref{Fform}), (\ref{rhoeq}) and (\ref{i1}), we get
\begin{equation}
\label{e1}
{W\over N\Ld^{3}}=
-{\partial_{\sigma}{\cal F}\over\Ld^{3}} = \int_{a}^{b}
{\d\varphi\over\pi}\,{ V(\varphi)\over\sqrt{(\varphi -a)(b-\varphi)}}
- 2\sigma\ln {b-a\over 4}\,\cdotp
\end{equation}
Then, by using (\ref{i2}), we finally obtain our fundamental matrix 
model equation
\begin{equation}
\label{fundmat}
\partial_{S} W (S,\L^{2},g_{p}) = -\ln\left( {b-a\over 4}\right)^{2N} .
\end{equation}
This equation shows in particular that the condition for a critical point, 
that yields $\langle S\rangle$, is simply $b-a = 4$. 

\subsection{Field theory and matrix model results are equivalent}

Comparison between (\ref{dW}) and (\ref{fundmat}) immediately yields the 
relationship between field theory and matrix model variables,
\begin{equation}
\label{r1}
{(b-a)^{2}\over 16} = {\Delta\over\L^{2}}\,\cdotp
\end{equation}
It remains to check that the equations (\ref{Acex}) and (\ref{Bcex}) that 
determine $\Delta$ are equivalent to the equations (\ref{m1}) and
(\ref{m2}) that determine $b-a$. The simplest way to do that is to work 
out explicitly the integrals in (\ref{m1}) and (\ref{m2}). This is very 
elementary. One changes the variable from $\varphi$ to $\psi =
2\left(\strut\varphi - (a+b)/2\right)/(b-a)$, expands the polynomial
$V'$ as a power series in $\psi$, and uses the identity
\begin{equation}
\label{intf}
\int_{-1}^{1}{\d\psi\over 2\pi}\, {\psi^{2p}\over\sqrt{1-\psi^{2}}} = 
{\C_{2p}^{p}\over 2^{2p+1}}\,\cdotp
\end{equation}
Formulas (\ref{m1}) and (\ref{m2}) are then exactly mapped onto 
(\ref{Acex}) and (\ref{Bcex}) respectively, provided one identifies
\begin{equation}
\label{r2}
{1\over 2} (a+b) = {z\over\L}\,\cdotp
\end{equation}
This completes the proof.

\section{Conclusion}

The non-trivial part of the field theory calculation made in Section 2
is the Intriligator-Leigh-Seiberg non-renormalization hypothesis
\cite{ILS}, while the non-trivial part of the matrix model calculation
made in Section 3 is the assumption that only planar diagrams
contribute \cite{DV}. In the model studied in the present paper, those
two hypothesis turn out to be equivalent. The fact that the linearity
in the couplings $g_{p}$ is implemented in the large $n$ matrix model
was a priori non-trivial. It is remarkable that the planar matrix
model formulas (\ref{m1}) and (\ref{m2}), which are manifestly linear,
could be identified with the integrating in relation (\ref{Bc}) for
$S$ and the integrating out relation (\ref{dzw}) for $z$, with the
suitable mapping between field theory and matrix model variables
(\ref{r1}) and (\ref{r2}). Linearity would be violated by
non-planar contributions. More generally, we conjecture that
the ILS linearity principle can be deduced from corresponding
linearity properties of the sum over {\it planar} diagrams.

There are two obvious directions of research that open up. The first 
is to try to generalize the approach of the present paper. It should 
not be too difficult, for example, to study the most general vacua of 
the theory (\ref{lag}), in particular by using the results of \cite{CV}. 
From the matrix model point of view, this amounts to generalizing 
equations like (\ref{fundmat}) to the multi-cut solutions. The second
is to try to use the exact superpotentials to work out some new 
interesting physics. For example, we have shown in \cite{fer} that
(\ref{ww3}) has several unexpected consequences. One can also use
the Dijkgraaf-Vafa proposal to full power, for theories that are not 
simple perturbations of ${\cal N}=2$. The interesting case of a 
Leigh-Strassler deformation of ${\cal N}=4$ has been treated in 
\cite{DH}. More generally, for pure ${\cal N}=1$ models,
the holomorphic variables that enter 
the DV superpotential are not physical, RG invariant, 
quantities. Nevertheless, the superpotential itself is physical, and 
it should contain some very interesting information. We are
presently investigating a two-matrix model \cite{2M} of this type.

\begin{appendix}
\section*{Appendix}
We wish first to derive equation (\ref{i1}),
\begin{equation}
\label{appeq}
I(\psi)=\int_{a}^{b} \d\varphi\, {\ln |\varphi -\psi|\over
\sqrt{(\varphi -a)(b-\varphi)}} =\pi \ln {b-a\over 4}\quad
{\rm for\ any}\ \psi\in [a,b]\, .
\end{equation}
The derivative of $I(\psi)$ can be expressed as
\begin{equation}
\label{a1}
{\d I\over\d\psi} = {1\over 2} \left(\strut f(\psi + i\epsilon) + f(\psi 
-i\epsilon)\right)
\end{equation}
for
\begin{equation}
\label{a2}
f(\psi) = \int_{a}^{b} {\d\varphi\over (\psi -\varphi)
\sqrt{(\varphi -a)(b-\varphi)}}\, \cdotp
\end{equation}
From 
\begin{equation}
\label{a3}
{1\over 2i\pi} \left(\strut f(\psi + i\epsilon) - f(\psi 
-i\epsilon)\right) = -{1\over\sqrt{(\psi -a)(b-\psi)}}
\end{equation}
and
\begin{equation}
\label{a4}
f(\psi) \mathrel{\mathop{\kern 0pt\sim}\limits_{\psi\rightarrow\infty}^{}}
{1\over\psi}\int_{a}^{b} 
{\d\varphi\over\sqrt{(\varphi -a)(b-\varphi)}} = {\pi\over\psi}\,\cvp
\end{equation}
we deduce
\begin{equation}
\label{a5}
f(\psi) = {\pi\over\sqrt{(\psi -a)(\psi -b)}}\,\cdotp
\end{equation}
The relation (\ref{a1}) then implies
\begin{equation}
\label{a6}
{\d I\over\d\psi} =\left\{\matrix{0&{\rm for\ }\psi\in [a,b]\, ,\cr
f(\psi) & {\rm for\ } \psi\not\in [a,b]\, .\cr}\right.
\end{equation}
It is then straightforward to compute $I (\psi)$ for $\psi>b$ by 
integrating 
(\ref{a6}) and using the asymptotics at infinity
$I (\psi) = \pi\ln |\psi| + {\cal 
O}(1/\psi)$. The formula (\ref{appeq}) follows from continuity.

We now turn to the derivation of equation (\ref{i2}). We have
\begin{equation}
\label{deri2}
\partial_{\sigma}\int_{a}^{b} {\d\varphi\over\pi}\, 
{ V(\varphi)\over\sqrt{(\varphi -a)(b-\varphi)}} =
{\partial a\over\partial\sigma}\oint_{\gamma} {\d\varphi\over 
4i\pi}\, {V(\varphi)\over (\varphi-a)^{3/2} (\varphi -b)^{1/2}} + 
(a\rightarrow b)\, ,
\end{equation}
where $\gamma$ is a contour encircling the cut $[a,b]$ 
counterclockwise. We then use
\begin{equation}
\label{inter1}
0 =\oint_{\gamma} \d\Bigl[ 
{ V(\varphi)\over\sqrt{(\varphi -a)(\varphi -b)}}\Bigr]
\end{equation}
and the condition (\ref{m2}) to deduce
\begin{equation}
\label{inter2}
\oint_{\gamma} {\d\varphi\, V(\varphi)\over (\varphi -a)^{3/2} 
(\varphi -b)^{1/2}} = - \oint_{\gamma} {\d\varphi\,
V(\varphi)\over (\varphi -a)^{1/2} (\varphi -b)^{3/2}}\,\cvp
\end{equation}
and we use
\begin{equation}
\label{inter3}
0 =\oint_{\gamma} \d\Bigl[ 
{\varphi V(\varphi)\over\sqrt{(\varphi -a)(\varphi -b)}}\Bigr]\, ,
\end{equation}
together with (\ref{m1}) and (\ref{inter2}), to deduce
\begin{equation}
\label{inter4}
\oint_{\gamma} {\d\varphi\over 2i\pi}\, 
{V(\varphi)\over (\varphi -a)^{3/2} 
(\varphi -b)^{1/2}} = {4\sigma\over a-b}\,\cdotp
\end{equation}
Equation (\ref{i2}) then follows immediately from (\ref{inter4}), 
(\ref{inter2}) and (\ref{deri2}).

\end{appendix}

\subsection*{Acknowledgements}

I am particularly indebted to Ken Intriligator. Without his insistence
on the generality of the linearity principle, and his patient
explanations of his work with F.~Cachazo and C.~Vafa \cite{CIV}, the
present paper would not have been written. I would also like to thank
J.-P.~Derendinger, R.~Hern\'andez, B.~Pioline, K.~Sfetsos, C.~Vafa and
particularly T.~Hollowood for useful discussions and/or
correspondences. This work was supported in part by the Swiss National
Science Foundation.


\begin{thebibliography}{99}
%
\bibitem{DV}{R.~Dijkgraaf and C.~Vafa, {\it A Perturbative Window into 
Non-Perturbative Physics,} HUTP-02/A034, ITFA-2002-34, hep-th/0208048.}
%
\bibitem{GV}{R.~Gopakumar and C.~Vafa, \atmp{3}{1999}{1415},\\
C.~Vafa, \jmp{42}{2001}{2798}.}
%
\bibitem{CIV}{F.~Cachazo, K.~Intriligator and C.~Vafa, 
\npb{603}{2001}{3}.}
%
\bibitem{DV1}{R.~Dijkgraaf and C.~Vafa, {\it Matrix Models, 
Topological Strings, and Supersymmetric Gauge Theories,} 
HUTP-02/A028, ITFA-2002-22, hep-th/0206255,\\
R.~Dijkgraaf and C.~Vafa, {\it On Geometry and Matrix Models,}
HUTP-02/A030, ITFA-2002-24, hep-th/0207106.}
%
\bibitem{CV}{F.~Cachazo and C.~Vafa, {\it ${\cal N}=1$ and ${\cal 
N}=2$ Geometry from Fluxes,} HUTP-02/A021, hep-th/0206017.}
%
\bibitem{SW1}{N.~Seiberg and E.~Witten, \npb{426}{1994}{19}, erratum
{\bf B 430} (1994) 485,\\
N.~Seiberg and E.~Witten, \npb{431}{1994}{484}.}
%
\bibitem{SW2}{P.~C.~Argyres and A.~E.~Faraggi,
\prl{\bf 74}{1995}{3931},\\
A.~Klemm, W.~Lerche, S.~Yankielowicz and S.~Theisen,
\plb{344}{1995}{169}.}
%
\bibitem{ILS}{K.~Intriligator, R.G.~Leigh and N.~Seiberg, 
\prd{50}{1994}{1092}.}
%
\bibitem{F}{F.~Ferrari, \npb{612}{2001}{151},\\
F.~Ferrari, \npb{617}{2001}{348},\\
F.~Ferrari, {\it Four dimensional non-critical strings,}
Les Houches summer school 2001, Session LXXVI,
{\it l'Unit\'e de la Phy\-si\-que fondamentale: Gravit\'e, Th\'eorie
de Jauge et Cordes,} A.~Bilal, F.~David, M.~Douglas and N.~Nekrasov 
editors, hep-th/0205171.}
%
\bibitem{fer}{F.~Ferrari, {\it Quantum parameter space and double 
scaling limits in ${\cal N}=1$ super Yang-Mills theory,}
NEIP-02-008, LPTENS-02/49, hep-th/0211069.}
%
\bibitem{sei1}{N.~Seiberg, \plb{318}{1993}{469}.}
%
\bibitem{revN1}{K.~Intriligator and N.~Seiberg, {\it Lectures on 
supersymmetric gauge theories and electric-magnetic duality,} 
\npps{55}{1997}{1}, hep-th/9509066.}
%
\bibitem{DH}{N.~Dorey, T.J.~Hollowood, S.P.~Kumar and 
A.~Sinkovics, {\it Exact Superpotentials form Matrix Models,} 
SWAT-350, hep-th/0209089,\\
N.~Dorey, T.J.~Hollowood, S.P.~Kumar and 
A.~Sinkovics, {\it Massive Vacua of ${\cal N}=1^{*}$ Theory and 
$S$-duality from Matrix Models,} SWAT-352, hep-th/0209099.}
%
\bibitem{ken}{K.~Intriligator, \plb{336}{1994}{409}.}
%
\bibitem{SV}{M.A.~Shifman and A.I. Vainshtein, \npb{277}{1986}{456},\\
M.A.~Shifman and A.I. Vainshtein, \npb{359}{1991}{571}.}
%
\bibitem{RR}{N.~Arkani-Hamed and H.~Murayama, \prd{57}{1998}{6638},\\
N.~Arkani-Hamed and H.~Murayama, \jhep{06}{2000}{30},\\
M.A.~Shifman, {\it Exact results in gauge theories: putting 
supersymmetry to work,} the 1999 Sakurai Prize lecture, 
\ijmpa{14}{1999}{5017}.}
%
\bibitem{RGN2}{M.~Matone, \plb{357}{1995}{342},\\
T.~Eguchi and S.K.~Yang, \mpla{11}{1996}{131},\\
J.~Sonnenschein, S.~Theisen and S.~Yankielowicz, \plb{367}{1996}{145}.}
%
\bibitem{DS}{M.R.~Douglas and S.H.~Shenker, \npb{447}{1995}{271}.}
%
\bibitem{matrev}{P.~Di Francesco, P.~Ginsparg and J.~Zinn-Justin,
\pr{254}{1995}{1}.}
%
\bibitem{BIPZ}{\'E.~Br\'ezin, C.~Itzykson, G.~Parisi and J.-B.~Zuber,
\cmp{59}{1978}{35}.}
%
\bibitem{eynard}{B.~Eynard, {\it Gravitation quantique 
bidimensionnelle et matrices al\'eatoires,} PhD thesis, University of 
Paris VI, 1995.}
%
\bibitem{2M}{F.~Ferrari and R.~Hern\'andez, in preparation.}
%

%
\end{thebibliography}
\end{document}